\documentclass[]{aa}
\usepackage{graphicx}
\begin{document}
\title{Photometric survey of marginally investigated open clusters: I. Basel~11b, King~14, Czernik~43 
\thanks{Based on observations at the Leopold-Figl Observatory for Astrophysics, University of Vienna.}}
\author{M.~Netopil\inst{1}, H.M.~Maitzen\inst{1}, E.~Paunzen\inst{1}, A.~Claret\inst{2}}

\mail{martin.netopil@univie.ac.at}

\institute{Institut f\"ur Astronomie der Universit\"at Wien,
           T\"urkenschanzstr. 17, A-1180 Wien, Austria
\and	   Instituto de Astrof\'isica de Andaluc\'ia
		   CSIC, Apartado 3004, 18080 Granada, Spain		   
}

\date{Received 2005; Accepted 2006}
\authorrunning{M. Netopil et al.}{}
\titlerunning{Survey of marginally investigated open clusters I.}{}

\abstract
{To progress in galactic studies based on open clusters, e.g. 
cluster formation rate or kinematical properties, one needs to 
improve the number of open clusters observed. However, only 
half of the 1700 known galactic open clusters have been properly 
observed so far, making any statistical investigation insignificant,
especially at larger distances from the Sun.
We study marginally investigated or 
neglected open clusters with Bessell CCD $BVR$ photometry, 
whose data were used to fit isochrones to the individual color-magnitude diagrams.
We examined the galactic clusters Basel~11b, King~14 and Czernik~43, 
the last being observed for the first time to this extent. As well as a careful 
comparison to available photometry, their parameters such as age, interstellar 
reddening, distance and apparent diameter were determined. The 
obtained cluster properties were verified by near infrared 2MASS data.
The three investigated intermediate age clusters are all located in the galactic
disk with distances between 1.8 and 3.0\,kpc from the Sun.

}
\maketitle

\section{Introduction}

In the comprehensive database for open clusters WEBDA (http://www.univie.ac.at/webda) there is a considerable lack of photometric data for about half of the more than 1700 known or suspected open clusters and therefore information about their age, distance and interstellar reddening is missing. Larger sets of open cluster parameters would be very useful to study their formation rate or life time.
After the first paper, by Netopil et al. (2005), with combined photometry in the $\Delta a$ and $BVR$ system of \object{NGC 7296} we have continued our efforts in this respect, to expand the target list for $\Delta a$ photometry using a filter system developed for the detection of chemically peculiar stars.
On the basis of WEBDA, the open clusters Basel~11b, King~14 and Czernik~43, all lying very close to  the galactic plane, were chosen for the ongoing survey in the Bessell $BVR$ system and compared to published $UBV$ and $RGU$ data as well as some individual $UBV$ measurements. The recent survey by Kharchenko et al. (2005a), hereafter K05a, based on a catalogue of stars in open cluster areas (Kharchenko et al. 2005b, K05b hereafter), mainly compiled and homogenised from Tycho data, includes also Basel~11b and Czernik~43. The latter was investigated for the first time in more detail and extent in the present work, although photometric measurements already exist. Cluster parameters like age, distance modulus and interstellar reddening were determined on the basis of two colour-magnitude diagrams each using isochrones by Claret (2004). For a further independent proof, the available 2MASS data and the corresponding $JHK_{S}$ isochrones by Bonatto et al. (2004) were examined, yielding excellent agreement for all three clusters. This data was also used to determine the cluster diameters by means of stellar density profiles. Of special interest are the immediate surroundings of King~14, including three known clusters, where at least two of them may represent a double cluster candidate as proposed by Subramaniam et al. (1995). 

\section{Observations and reduction}

The CCD observations with the Bessell filters $BVR$ of the open clusters Basel~11b, King~14 and Czernik~43 were obtained on 2004 November 25 at the 1.52-m telescope of the Leopold-Figl Observatory for Astrophysics (FOA) of the University of Vienna, located at Mt. Mittersch\"opfl (Austria) using the multimode instrument OEFOSC and a SITe SI502AB CCD as detector, covering a field-of-view of 5.75$\arcmin$ square (512$\times$512 px, 1\,px = 0.674\arcsec). The seeing conditions were between 1.6 and 2.2\arcsec. Since the CCD is equipped with a single-step Peltier cooling system, not ensuring a stable temperature 
during the whole observing night, we possibly experience zero point shifts between programme and standard objects, complicating a correct standard transformation. A comparison with published data was therefore helpful, although not straightforward in the case of marginally investigated open clusters.

The basic CCD reductions (bias-subtraction, dark-correction, flat-fielding) were carried out within standard IRAF V2.12.2 routines. For all frames we have applied a point-spread-function fitting within the IRAF task DAOPHOT (Stetson 1987). Photometry of each frame was performed separately and the measurements were then averaged and weighted by their individual photometric error.

In addition to the programme clusters we have observed the secondary standard field of \object{NGC 7790} by Stetson (2000) during the whole same night, used for the transformation to standard magnitudes and the determination of atmospheric extinction. After applying the aperture correction using several bright isolated stars, the transformation between standard and instrumental magnitudes was performed using the coefficients given in Table \ref{coef}. As already mentioned in Netopil et al. (2005) the actual Bessel $BVR$ filters seem to match the ``standard'' $UBV$ system quite well, so no colour dependency, except for the $B$-filter, was found. The typical errors for the individual $BVR$-photometry are up to 0.03\,mag and are also given in the tables provided electronically. The observing log with the number of frames in each filter, exposure times and airmass range is listed in Table \ref{Obslog}.

\begin{table}
\caption{The observing log with the number of total frames in each filter, exposure times in seconds and Airmass $X$.}
\label{Obslog}
\centering
\label{Obslog}
\begin{tabular}{lccc}
\hline\hline
 & Basel~11b & King~14 & Czernik~43\\
\hline
$B$ 	& 7 (200s) & 7 (50/200s) & 7 (200s) \\
$V$ 	& 8 (100s) & 13 (40/80/160s) & 11 (80/160s) \\
$R$ 	& 16 (30/50/100s) & 15 (10/30/60s) & 17 (10/30/60s) \\
$X$ 	& 1.14 - 1.31 & 1.07 - 1.15 & 1.03 - 1.08 \\
\hline
\end{tabular}
\end{table}

\begin{table}
\caption{The regression coefficients for the transformation from instrumental $bvr$ to standard $BVR$ magnitudes as well as the coefficients of the atmospheric extinction. The errors in the final digits of the corresponding quantity are given in parenthesis.}
\label{coef}
\centering
\begin{tabular}{cl}
\hline\hline
 & transformation coefficients\\
\hline
$B$     & $= b - 0.728(6) + 0.307(9)\cdot(B - V)$\\
$V$     & $= v - 0.237(2)$\\
$R$     & $= r + 0.097(3)$\\
\hline
$B$     & $: 0.226(9)\cdot X$\\
$V$ 	& $: 0.127(1)\cdot X$\\
$R$ 	& $: 0.103(9)\cdot X$\\
\hline
\end{tabular}
\end{table}

\section{Results} \label{results}

\begin{figure*}
\begin{center}
\includegraphics[width=180mm]{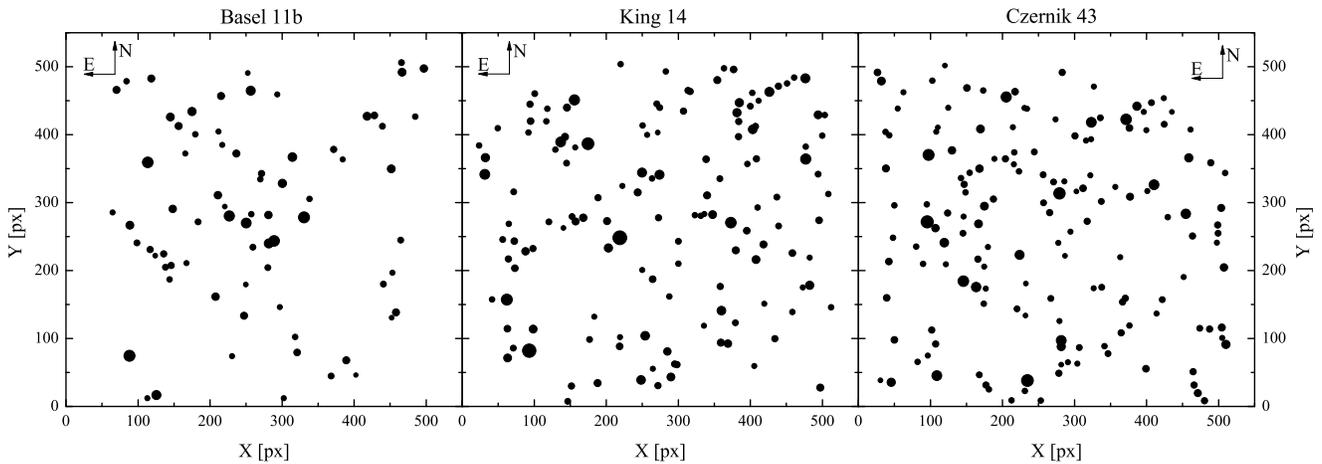}
\caption[]{The observed fields of Basel~11b, King~14 and Czernik~43 (1\,px = 0.674\arcsec).}
\label{charts}
\end{center}
\end{figure*}

\begin{figure*}
\begin{center}
\includegraphics[width=160mm]{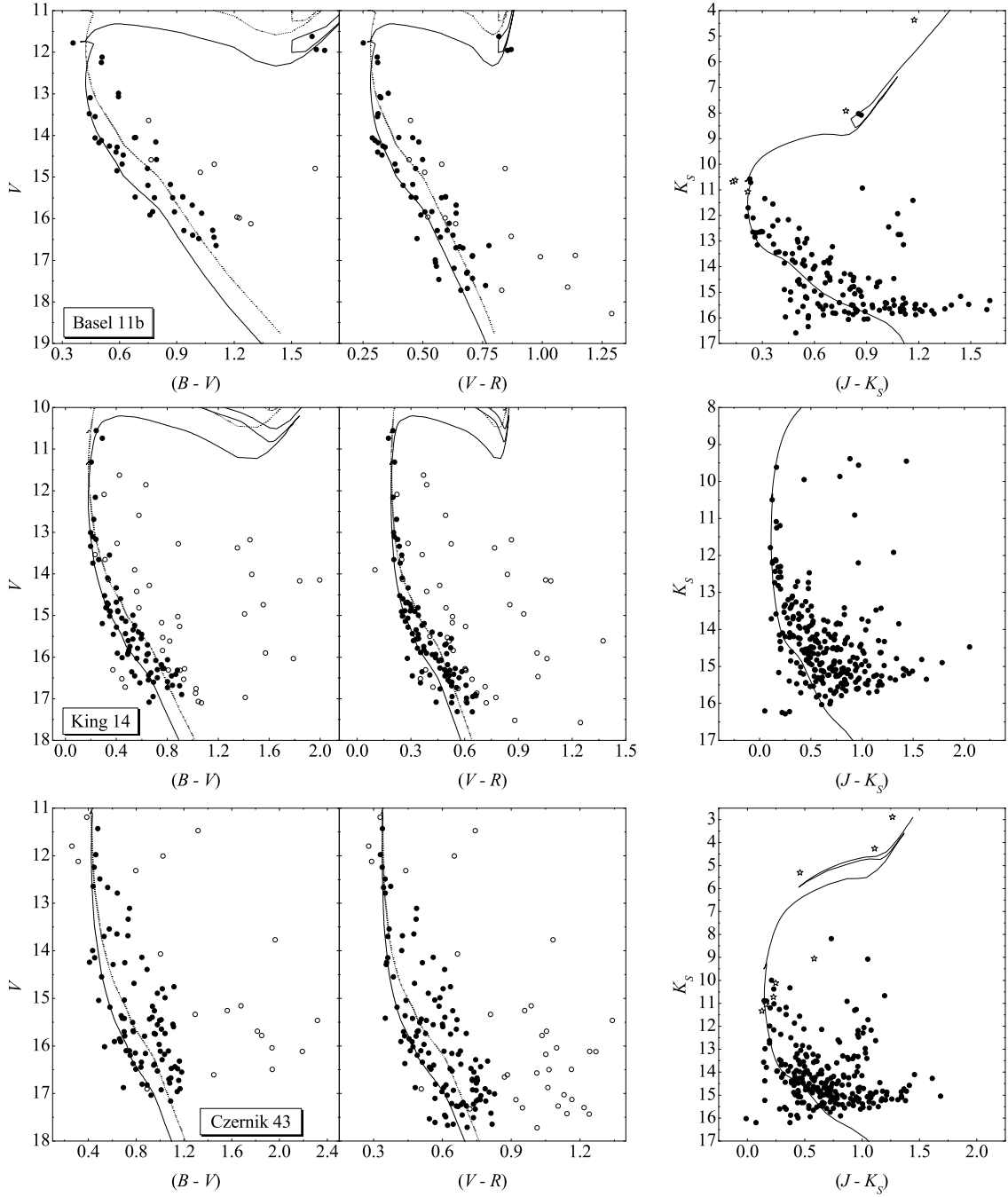}
\caption[]{The colour-magnitude diagrams for the programme clusters Basel~11b, King~14 and Czernik ~43. For the $BVR$ photometry the isochrones by Claret (2004), for the 2MASS data the appropriate isochrones by Bonatto et al. (2004) were used to determine the cluster parameters at solar metallicity, which are listed in Table \ref{results}. The dotted lines represent the upper binary ridge (isochrone shifted by 0{\fm}75), the open circles probable nonmembers mainly based on literature as well as the presented colour-magnitude diagrams. To minimize the contamination by field stars in the 2MASS colour-magnitude diagrams, only stars lying within a diameter of 5$\arcmin$ were used. Stars, designated as probable members due to their proper motion by K05b are marked by open asterisks and were also included, even if lying outside the above mentioned diameter.}
\label{cmd}
\end{center}
\end{figure*}

\begin{table}
\caption{Summary of the results for the observed programme clusters. The distances resulted from the averages of the determined distance moduli, and the cluster coordinates using the radial density profiles. The errors in the final digits of the corresponding quantity are given in parenthesis. Parameters indicated by (*) are based on the photometry by Pandey et al. (2001).}
\label{results}
\centering
\begin{tabular}{lccc}
\hline\hline
 & Basel~11b & King~14 & Czernik~43\\
 & \object{C 0555+219} & \object{C 0029+628} & \object{C 2323+610}\\
\hline
$l/b$					&	187.44/$-$1.11&	120.73/+.38	& 112.84/+0.17	\\
Ra (2000)				&	05 58 10&	00 32 03& 23 25 47			\\
Dec	(2000)				&	+21 57 50&	+63 09 20& +61 21 30	\\
log\,$t$				&	8.4(1)	&	7.9(1)	&	7.6(1)	\\
$m_{V}-M_{V}$ 			&	12.7(2)	&	13.4(2)&	13.9(2)	\\
$m_{K_{S}}-M_{K_{S}}$ 	&	11.5(2) &	12.5(2) &	12.2(2)	\\
$E(B-V)$  				&	0.48(5)	&	0.34(5)	&	0.62(5)	\\
$E(V-R)$ 				&	0.31(5)	&	0.26(5)	&	0.42(5)	\\
$E(V-I)$ 				&	     	&	    	&	0.85(5)*\\
$E(U-B)$ 				&	    	&	    	&	0.38(5)*\\
$E(J-K_{S})$ 			&	0.25(3)	&	0.21(3)	&	0.28(3)	\\
d [kpc] 				&	1.80(25)&	2.96(42)&	2.51(36)\\
$R_{GC}$\,[kpc]			&	10.29(25)&	10.33(29)&	9.75(22)\\
$|z|$\,[pc]				&	35(5)  &	20(2)  &	7(1)   \\
diameter [$\arcmin$]	&	7		&	8		&	5		\\
\# stars				&	73		&	136		&	154		\\
\hline
\end{tabular}
\end{table}

In the following we will discuss the results and the comparison to
the literature for the individual open clusters in more detail. The stars are numbered in order of ascending (X,Y) coordinates
on the CCD frames. 

To determine the cluster parameters, isochrones by Claret (2004) for the $BVR$ photometry and the corresponding ones for 2MASS by Bonatto et al. (2004) for solar metallicity were used. Since the affect of different metallicities on the isochrone fitting procedure was already discussed by Pinsonneault et al. (2004), the assumption of solar abundances can be justified, if inspecting the determined ages and galactocentric distances of the clusters listed in Table \ref{results}, and from the work of Chen et al. (2003), who found a relative low radial iron gradient for the Milky Way of $-$0.024 $\pm$ 0.012 dex kpc$^{-1}$ for clusters younger than 0.8\,Gyr. 

To be able to compare the results of the $BVR$ and 2MASS photometry, the relations $E(J-K_{S})=0.488\,E(B-V)$ by Bonatto et al. (2004) and $A_{K_{S}}=0.67\,E(J-K_{S})$ by Dutra et al. (2002) were used. 

The listed differences of our photometry compared to the literature values are always given as $\Delta$ (literature $-$ present data) of the corresponding quantity. 

To obtain an estimate of the cluster diameters, the stellar density profiles (Fig. \ref{density}) were prepared using 2MASS data, which cover a much larger area, and the King (1962) profile. The star density was derived by conducting star counts in concentric rings up to 10$\arcmin$ around the center given in Table \ref{results}, divided by their respective area, whereas the average of the outer five arcminutes plus the standard deviation was used as the separation from the galactic field.  
\subsection{Basel~11b}
The open cluster Basel~11b was investigated by Grubissich (1973) using photographic plates in the $RGU$ photometric system. The author obtained a true distance modulus of 10$\fm$88 (1500\,pc) and an interstellar reddening of $E(G-R)$=1.01. Using the relation $E(G-R)$=1.39$E(B-V)$ (Steinlin 1968), the corresponding $E(B-V)$ is 0.73, much higher than our result of 0.48 (Table \ref{results}). 
Our investigation has 39 objects in common with that study, which were used for a photometric comparison. In order to transform the $RGU$ measurements into the $UBV$ system, the calibrations by Buser (1978) were taken 
\begin{eqnarray*}
V = G -0.495(G-R)-0.008(U-G)+0.179+\alpha \\ 
(B-V) = +0.784(G-R)+0.027(U-G)-0.302+\alpha
\end{eqnarray*}
with $\alpha$ depending on the colour and interstellar reddening. For the latter the value of Grubissich (1973) was used, since the differences of the final calibrated photometry are of the order of 0{\fm}01, if one uses an $E(B-V)$ of 0.73 or 0.48 as obtained in this work. The photometric comparison, which can be found in Figure \ref{comparison}, yields $\Delta V$=+0.06 and $\Delta (B-V)$=+0.07, whereas nonmembers according to Grubissich (1973) were rejected for the calculation.  
However, it is difficult to derive a zeropoint shift of our photometry, since one has to take into account the accuracy of photographic plates and the errors introduced by the transformation. 
We have determined the cluster parameters using the isochrones by Claret (2004) resulting in a best fit of log\,$t$\,=\,8.4\,$\pm$\,0.1, $m_{V}-M_{V}$\,=\,12.7\,$\pm$\,0.2, $E(B-V)$\,=\,0.48\,$\pm$\,0.05 and $E(V-R)$\,=\,0.31\,$\pm$\,0.05. Using near infrared 2MASS data and fitting isochrones from Bonatto et al. (2004), we get the result shown in Figure \ref{cmd}. The isochrone fitting procedure was performed using the parameters determined via $BVR$ photometry as an initial estimation, varying them to obtain the best possible fit to the available data.     
This open cluster is also included in the catalogue by K05a, who determined log\,$t$\,=\,8.82\, $m_{V}-M_{V}$\,=\,11.19 and $E(B-V)$\,=\,0.10 on the basis of three stars. These parameters are not in agreement with our results. However, our isochrone rests on the assumption that the three evolved stars (\#8, 46 and 54) in Figure \ref{cmd} are members due to their position in the  colour-magnitude diagrams, as well as the fact that at least two of them are lying close to the cluster core and that star \#54 is a probable member according to K05b, based on photometry, position and proper motion. The colour indices $(B-V)_{0}$, $(V-R)_{0}$ as well as the dereddened near infrared colours by 2MASS determine all these stars as of spectral type K0\,II/III. To be able to compare the 2MASS colours of these stars to the spectral type relation by Bessell \& Brett (1988), they were transformed to their homogenised system using the transformation by Carpenter (2001). The intrinsic colours and absolute magnitudes of the probable red giants are listed in Table \ref{giants}. Since stars \#8 and 54 were also measured by Grubissich (1973), who determined them as red giants, but nonmembers, the status of these stars has to be clarified by further spectroscopic investigations. If they should turn out to be nonmembers, the cluster would be much younger. On the basis of the stellar density profile (Fig. \ref{density}), the apparent diameter of Basel~11b was determined to 7 arcminutes, slightly lower than the value by Grubissich (1973), who lists 10$\arcmin$.
\\
\begin{table}
\caption{The intrinsic colour indices and absolute magnitudes of the possible red giant members within Basel~11b point to a spectral type of about K0\,II/III. The star numbers are according to our numbering system, within parenthesis the corresponding WEBDA-number is given. The near infrared 2MASS colour was transformed to the homogenised system by Bessel \& Brett (1988) using the transformation given in Carpenter (2001), the values of the corresponding interstellar reddening are listed in Table \ref{results}.}
\label{giants}
\centering
\begin{tabular}{lccccc}
\hline\hline
Star & $(B-V)_{0}$ & $(V-R)_{0}$ & $(J-K)_{0}$ & $M_{V}$\\
\hline
8(27) & +1.08 & +0.52 & +0.65 & $-$0.97\\
46     & +1.12 & +0.51 & +0.63 & $-$0.95\\
54(3)  & +1.06 & +0.47 & +0.56 & $-$1.28\\
\hline
\end{tabular}
\end{table}

\subsection{King~14}
\begin{figure}
\begin{center}
\includegraphics[width=85mm]{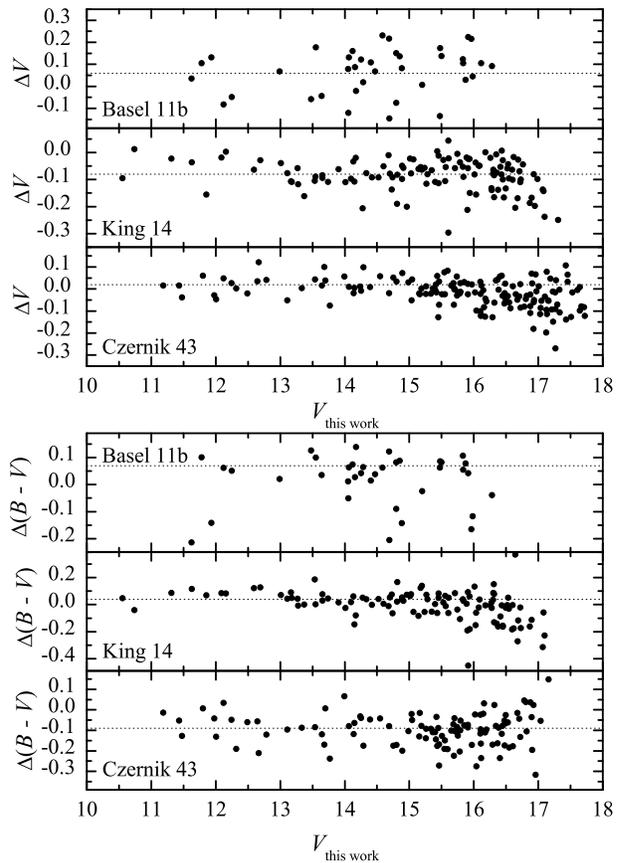}
\caption[]{Comparison of our photometry with literature values for the programme clusters. The dotted lines correspond to the determined offsets. The $RGU$ measurements of Basel~11b were transformed to the $UBV$ system as given in Section \ref{results}. Only stars down to 15{\fm}0 were used, and outliers were rejected for the zeropoint determinations. The offsets of our photometry (literature $-$ present work) are therefore $\Delta V$=+0.06, $\Delta (B-V)$=+0.07 for Basel~11b, $\Delta V$=$-$0.08, $\Delta (B-V)$=+0.04 for King~14 and $\Delta V$=+0.02, $\Delta (B-V)$=$-$0.09 for Czernik~43 respectively.}
\label{comparison}
\end{center}
\end{figure}

The open cluster King~14 was investigated photographically by Hardorp (1960) and Jasevicius (1964). Since the photometry of both studies are in agreement within the limits of photographic accuracy, the first reference was used for comparison (Figure \ref{comparison}) of the measurements, because it offers more stars in common within a broader colour range. We determined offsets as $\Delta V$=$-$0.08 and $\Delta (B-V)$=+0.04. The former photometry was carried out using $RGU$ plates, but its results were already transformed to the $UBV$ system by the author. The star \#93 (\#133 according to WEBDA) was measured  photoelectrically by Haug (1970) and Sarg \& Wramdemark (1977) in perfect concordance, yielding zeropoints for our photometry of $\Delta V$=$-$0.08 and $\Delta (B-V)$=+0.04, which are identical to the previous offsets.
The determined cluster parameters log\,$t$\,=\,7.9\,$\pm$\,0.1, $m_{V}-M_{V}$\,=\,13.4\,$\pm$\,0.2, $E(B-V)$\,=\,0.34\,$\pm$\,0.05 and $E(V-R)$\,=\,0.26\,$\pm$\,0.05 are comparable to those listed in the catalogue by Dias et al. (2002) and deviate somewhat from those of Hardorp (1960), who lists $E(B-V)$\,=\,0.47 and $m_{V}-M_{V}$\,=\,13.7$\pm$0.3. Examination of the 2MASS photometry using appropriate isochrones (Figure \ref{cmd}) strengthens our findings. Using the stellar density profile (Figure \ref{density}), we determined a cluster diameter of about 8 arcminutes, whereas the catalogue by Dias et al. (2002) lists the somewhat lower value of 6$\arcmin$. Unfortunately, this cluster is not included in the survey by K05a. In their ASCC-2.5 catalogue of stars in open cluster areas (K05b) all included stars in the surroundings of King~14 (the five brightest stars in Figure \ref{cmd}) are listed with member probabilities of zero. Since the cluster has a well defined main sequence, this situation seems unlikely. But, they have associated these stars with \object{NGC 146}, an open cluster lying about 12$\arcmin$ northeast of King~14. Using the cluster proper motion of King~14 by Glushkova et al. (1997) and the individual proper motions in the ASCC-2.5 catalogue, positive membership results for at least the brightest star in our data set (\#48, \#93 according to WEBDA). These two clusters are candidates for a cluster pair according to Subramaniam et al. (1995). Recently, the parameters for NGC~146 were estimated to about 10$-$16\,Myr, 3470\,pc and $E(B-V)=0.55$, respectively, by Subramaniam et al. (2005). They supposed such a young age for King~14, as they found one star showing H$\alpha$ emission using slitless spectral observations of this cluster. However, the age of about 80\,Myr determined for King~14 has to be considered as an upper limit, since it is difficult to fit an isochrone without a giant branch. Together with a loose concentration of stars, belonging to \object{NGC 133}, lying also very close by, the abovementioned clusters form an apparent triplet, although the latter is much nearer (630\,pc according to Carraro 2002). The author has interpreted his result as strong contamination of field stars. Since the interstellar reddening $E(B-V)=0.6$ is in a comparable range to the other clusters, which are much more distant, this result has to be revised in line with a homogeneous analysis of the whole region to clarify their association. Except for the nearly identical proper motions of the clusters King~14 and NGC~146 by Glushkova et al. (1997), the currently available information for them, especially the distances, does not support a physical relation.          
\\ 

\subsection{Czernik~43}
\begin{figure}
\begin{center}
\includegraphics[width=85mm]{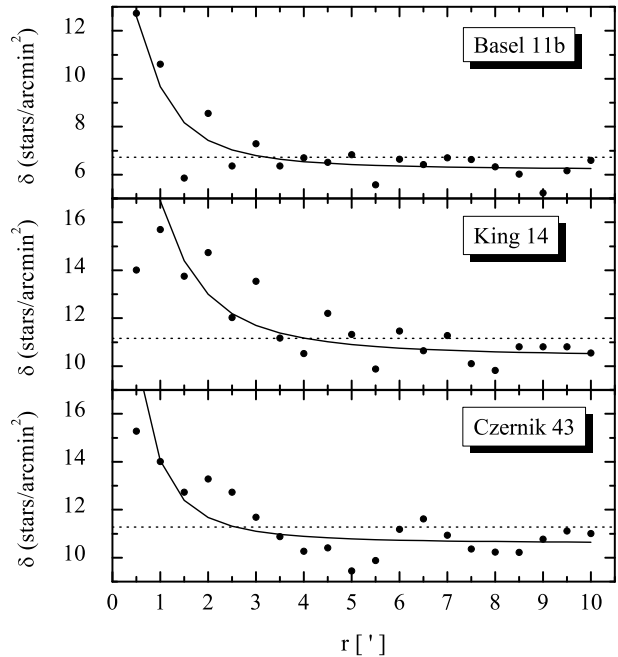}
\caption[]{The stellar density profiles of the programme clusters on the basis of the 2MASS data, using concentric rings around the determined cluster center given in Table \ref{results}. The dotted lines correspond to the average of the outer five arcminutes plus the standard deviation of the star density of that area ($\sigma_{\delta} \sim 0.5$ for all clusters) and the solid lines represent the best fit to the empirical King (1962) model. The cluster diameter estimates are therefore 7$\arcmin$, 8$\arcmin$ and 5$\arcmin$ for Basel~11b, King~14 and Czernik~43 respectively.}
\label{density}
\end{center}
\end{figure}

\begin{figure}
\begin{center}
\includegraphics[width=90mm]{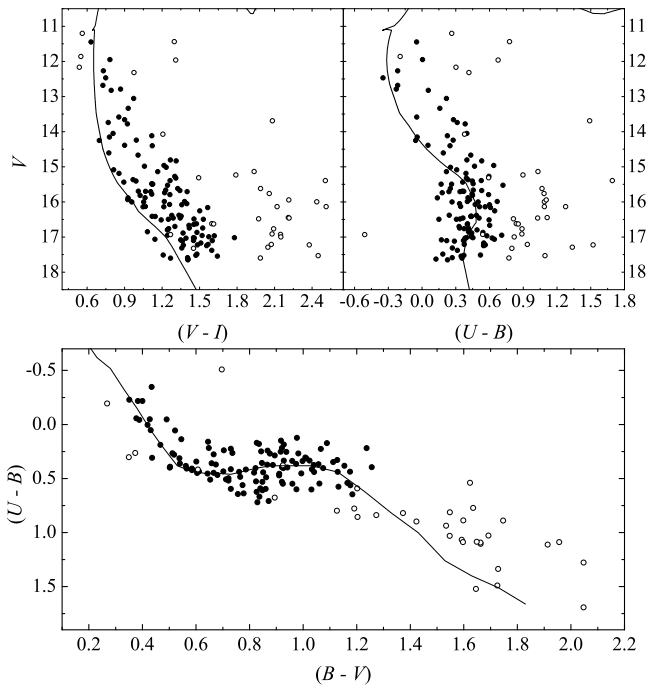}
\caption[]{The remaining colour-magnitude and colour-colour diagrams for Czernik~43 based on the photometry by Pandey et al. (2001). Isochrones by Claret (2004) were used to fit the data. Open circles represents probable nonmembers according to the available photometry, but due to the large scatter within the $(B-V)/(U-B)$ diagram only a preliminary membership analysis is possible. The best isochrone fit for log\,$t$\,=\,7.6 and $m_{V}-M_{V}$\,=\,13.9 yields $E(V-I)$\,=\,0.85\,$\pm$\,0.05 and  $E(U-B)$\,=\,0.38\,$\pm$\,0.05. The solid line within the $(B-V)/(U-B)$ plane indicates 
the Schmidt-Kaler (1982) ZAMS shifted by $E(B-V)$\,=\,0.53 and $E(U-B)$\,=\,0.38.
}
\label{UBVI}
\end{center}
\end{figure}

The open cluster Czernik~43 was investigated photometrically for the first time, but K05a have included it in their survey using the all-sky compiled
catalogue of 2.5 million stars of galactic open cluster areas (K05b) to determine its parameters. They found log\,$t$\,=\,7.7, $E(B-V)$\,=\,0.7 and $m_{V}-M_{V}$\,=\,14.16, but only one star was used for the calculation. However, their results are in good agreement with that presented in this work (log\,$t$\,=\,7.6\,$\pm$\,0.1, $m_{V}-M_{V}$\,=\,13.9\,$\pm$\,0.2, $E(B-V)$\,=\,0.62\,$\pm$\,0.05 and  $E(V-R)$\,=\,0.42\,$\pm$\,0.05). Adding 2MASS data confirms this finding (Figure \ref{cmd}, Table \ref{results}). Some evolved stars (ASCC 151568, 151607 and 151652), probable members according to their proper motions (K05b), lie outside of our field-of-view, but were included in the 2MASS colour-magnitude diagram, resulting in an additional proof of our determined parameters. These stars have spectral types of K2, A5\,Ia and K2 respectively, according to K05b. One star (\#9, [H96] PW Cas 5) within our observed field, measured by Henden (1996) who used it as comparison star for a study of faint cepheids, yields differences to our photometry of $\Delta V$=$-$0.002, $\Delta (B-V)$=+0.074 and $\Delta (V-R)$=$-$0.064. However, it is difficult to derive a zeropoint on the basis of a single star. The determined cluster diameter of 5$\arcmin$ (Fig. \ref{density}) is comparable to the value of 6$\arcmin$ by Dias et al. (2002). Czernik~43 is located about 16$\arcmin$ away from the cluster \object{NGC 7654}, which was investigated by Pandey et al. (2001) using a Schmidt telescope with a large field-of-view, including also our programme cluster. Except for three stars, we were able to identify all our objects in their data set of nearly 18000 measured stars. However, Pandey et al. (2001) did not use their data to investigate Czernik~43 as a galactic cluster. Unfortunately, the above mentioned evolved stars were not measured by them, most likely because of their brightness. Using this photometry as comparison, we found $\Delta V$=+0.02 and $\Delta (B-V)$=$-$0.09 of our measurements, whereas $\Delta (B-V)$ deviates extremely to the determined offset on the basis of [H96] PW Cas 5 and the photometry by Henden (1996). 
To obtain further proof, the data by Pandey et al. (2001) were compared with available photoelectric and CCD photometry of the well investigated cluster NGC~7654 (Pesch 1960, Hoag et al. 1961, Choi et al. 1999, Stetson 2000) resulting in nearly constant zeropoint shifts to all references of $\Delta V$=$-$0.03, $\Delta (B-V)$=+0.02 and $\Delta (U-B)$=+0.02 (other photometry $-$ Pandey et al. 2001) for stars down to $\approx$ 15{\fm}0, that implies a somewhat lower offset of our photometry. The $\Delta (V-I)$ values deviate between +0.02 and $-$0.04 using the data by Stetson (2000) and Choi et al. (1999) respectively. For fainter stars a larger scattering and gradients can be found as also reported by Pandey et al. (2001). The results are comparable to their analysis, except the one to Stetson (2000), where they found much larger deviations (more than 0{\fm}1). Since the photometry of the standard fields by Stetson (2000) are available on the corresponding webpages only, which are updated regularly, a clarification of that deviation is not possible. The remaining colour-magnitude and colour-colour diagrams based on the photometry by Pandey et al. (2001) can be found in Fig. \ref{UBVI}. Due to the large scatter (especially of the $(U-B)$ colour), only a preliminary membership analysis is possible. However, the additional colour indices provide further evidence of our determined cluster parameters. The obtained difference in $E(B-V)$ (0{\fm}09) can be explained by the offset to the photometry of Pandey et al. (2001). They also state a differential reddening within NGC~7654 (mean $E(B-V)=0.57$), which can be assumed as well for Czernik~43 due to the width of the cluster main sequence.   

\section{Conclusions}
On the basis of CCD $BVR$-photometry for three open clusters (Basel~11b, King~14, Czernik~43), their parameters age, reddening and distance (as given in Table \ref{results}) were determined using the isochrones by Claret (2004). Besides a careful comparison with available measurements to get an idea of a possible offset of our photometry, the cluster properties were compared with the literature values, yielding divergent results, however the programme clusters have been only marginally investigated up to now. After analyzing the 2MASS data, we are able to confirm the determined parameters with excellent agreement. One open cluster was investigated by us for the first time in an appropriate way, although photometric CCD measurements have been provided by Pandey et al. (2001), who examined the nearby cluster NGC~7654, but ignoring the less conspicuous cluster Czernik~43. The question of a possible triple system, containing King~14, NGC 133 and NGC 146 was raised, but further homogeneous investigations are necessary to clarify their affiliation. For all three programme clusters a normal extinction law ($R_{V}$ = 3.1) can be assumed when inspecting the various values of interstellar reddening (Table \ref{results}).    
    
The tables with all the data of the observed stars are available in electronic form at the CDS (cdsarc.u-strasbg.fr), WEBDA (univie.ac.at/webda) or upon request from the first author. These tables include the coordinates ($X/Y$ within our frames, $\alpha$/$\delta$), $V$ magnitudes, the colour indices, the corresponding errors as well as a cross identification with the literature and WEBDA.

\begin{acknowledgements}
This research was performed within the project {\sl P17920} of the Austrian Fonds zur F{\"o}rderung der 
wissen\-schaft\-lichen Forschung (FwF) and benefitted also from the financial contributions of the City of Vienna (Hochschuljubil{\"a}umsstiftung projects: H-111/1995, H-1123/2002). 
M.~Netopil acknowledges the support by a ``Forschungsstipendium'' from the University of Vienna. 
Use was made of the SIMBAD database, operated at CDS, Strasbourg, France, 
the WEBDA database, operated at the University
of Vienna, Austria, NASA's Astrophysics Data System, and of data products from the Two Micron All Sky Survey. We would like to thank the anonymous referee for very useful comments.
\end{acknowledgements}

\end{document}